\documentstyle[here,epsfig]{mn2e}
\def\simlt{\mathrel{\rlap{\lower 3pt\hbox{$\sim$}}\raise 2.0pt\hbox{$<$}}}
\def\simgt{\mathrel{\rlap{\lower 3pt\hbox{$\sim$}} \raise 2.0pt\hbox{$>$}}}

\def\gtsima{$\; \buildrel > \over \sim \;$}\def\gtsima{$\; \buildrel > \over
  \sim \;$}

\def\ltsima{$\; \buildrel < \over \sim \;$}
\def\gtrsim{\lower.5ex\hbox{\gtsima}}
\def\lesssim{\lower.5ex\hbox{\ltsima}}

\newcommand{\q}{\begin{equation}}
\newcommand{\qa}{\begin{eqnarray}}
\newcommand{\qs}{\begin{eqnarray*}}
\newcommand{\nq}{\end{equation}}
\newcommand{\nqa}{\end{eqnarray}}
\newcommand{\nqs}{\end{eqnarray*}}

\begin{document}

\title[BSS: radial distribution and progenitors] 
{The radial distribution of blue straggler stars and the nature of their progenitors}

\author[M. Mapelli, S. Sigurdsson, F. R. Ferraro, M. Colpi,
  A. Possenti, B. Lanzoni]
{M. Mapelli$^{1}$, S. Sigurdsson$^{2}$, F. R. Ferraro$^{3}$,
M. Colpi$^{4}$, A. Possenti$^{5}$, B. Lanzoni$^{6}$
\\
$^1$S.I.S.S.A., Via Beirut 2 - 4, I--34014 Trieste, Italy; {\tt
mapelli@sissa.it}\\ $^2$Department of Astronomy and Astrophysics, The
Pennsylvania State University, 525 Davey Lab, University Park,
PA~16802\\ $^3$Dipartimento di Astronomia, Universit\`a di Bologna,
via Ranzani 1, I--40126 Bologna, Italy \\ $^4$Dipartimento di Fisica
G. Occhialini, Universit\`a di Milano Bicocca, Piazza della Scienza
3. I--20126 Milano, Italy\\ $^5$INAF, Osservatorio Astronomico di
Cagliari, Poggio dei Pini, Strada 54, I--09012 Capoterra, Italy\\
$^6$INAF, Osservatorio Astronomico di Bologna, via Ranzani 1, I--40126 
Bologna, Italy\\}

\maketitle 
\vspace {7cm}

\begin{abstract}
The origin of blue straggler stars (BSS) in globular clusters (GCs) is still not fully
understood: they can form from stellar collisions, or through mass transfer in isolated, primordial binaries (PBs).  
In this paper we use the radial distribution of BSS
observed in four
GCs (M3, 47~Tuc, NGC~6752 and $\omega{}$ Cen) to
investigate  which formation process prevails. We find 
that both channels co-exist in all the considered GCs. 
%They generate a similar spatial distribution
%both in the low density
%cluster M3 and in the much denser 47~Tuc and NGC~6752. 
The fraction of mass-transfer (collisional) BSS with respect to the total number of BSS is around 
$\sim 0.4-0.5$ ($\sim 0.5-0.6$) in M3, 47~Tuc, and NGC~6752.  
The case of $\omega$~Cen is
peculiar with an underproduction of collisional BSS. The relative lack of collisional BSS in $\omega$~Cen
can be understood if mass segregation has not yet driven to the core a
sizeable number of PBs, 
which dominate stellar collisions through three- and four-body processes.
The spatial distribution of BSS
provides strong hints to their origin: the BSS
in the cluster outskirts form almost exclusively from mass transfer in PBs, 
whereas the BSS found close to the cluster core
most likely have a collisional origin. 
%The contamination,  in the cluster periphery, of binaries ejected from collisions is found to be unimportant. 

%{\bf Also NGC 6752 is peculiar and its small fraction of blue
%stragglers today is likely to be related to its current 
%state of core-collapsed cluster.}
  
\end{abstract}


\section{Introduction}

Blue straggler stars (BSS) are stars lying above and blue-ward of the
turn-off in the color-magnitude diagram (CMD) of a star cluster.  At
least two different processes have been proposed to explain their
formation (Fusi Pecci et al. 1992; Bailyn 1995; Bailyn and Pinsonneault 1995; Procter Sills,
Bailyn \& Demarque 1995; Sills \&{} Bailyn 1999; Sills et al.  2000;
Hurley et al. 2001).  
The first, dubbed the {\it mass-transfer 
scenario}, suggests that BSS are generated by primordial binaries
(hereafter PBs) that evolve mainly in isolation until they start
mass-transfer and possibly coalesce (McCrea 1964; Carney et
al. 2001).  The second, known as the {\it collisional 
scenario}, states that BSS are the product of a merger between two
main sequence stars (MSs) in a dynamical interaction that involves a
MS-MS collision,  most likely in  a 
binary-MS encounter (Davies, Benz \& Hills
1994; Lombardi et al. 2002).  
The collisional BSS (COL-BSS) differ kinematically from the mass-transfer BSS (MT-BSS), since they are believed to acquire kicks
due to dynamical recoil. 
In both these hypotheses, the resulting BSS have  mass
exceeding the turn-off mass of the cluster and are fueled by hydrogen
thanks to the mixing of the hydrogen-rich surface layers of the two
progenitor stars.

%%%%%%%%%%%%%%%%%%%%%% FIGURE 1%%%%%%%%%%%%%%%%%%%%%%%%%%%%%%%%
\begin{figure}
\epsfxsize=8.truecm 
\epsfysize=8.truecm
\epsfbox{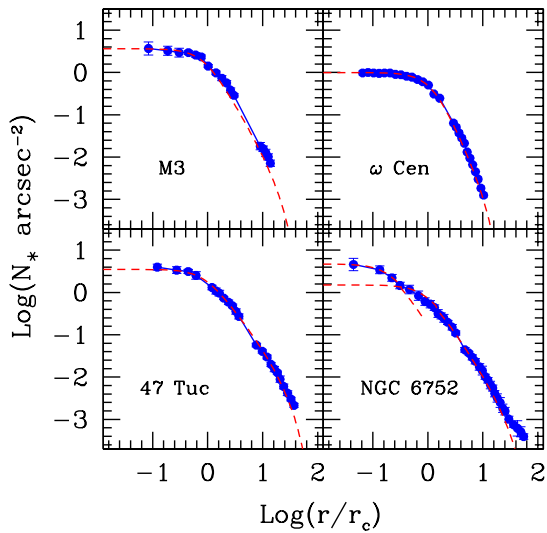}
\label{fig1}
\vspace{-0.5truecm}
\caption
{Comparison between the observed surface star density profile
of the considered GCs (solid line and full circles) and the adopted
multi-mass King model (dashed line). The units on the y-axis are
number of stars per arcsec$^2.$ The observed profiles are taken
from Ferraro et al. (1997a) for M3, Ferraro et al. (2006a) for
$\omega$~Cen, Mapelli et al. (2004) for 47~Tuc, Ferraro et al. (2003b)
for NGC~6752. }
\end{figure}
%%%%%%%%%%%%%%%%%%%%%%%%%%%%%%%%%%%%%%%%%%%%%%%%%%%%%%%%%%%%%%%%%%%%%%%%

The two aforementioned scenarios do not necessarily exclude each other
and might co-exist within the same star cluster (Ferraro et al. 1993; Davies, 
Piotto \& De Angeli 2004; Mapelli et al. 2004).  However, until recently, 
it has been difficult to
estimate the relative importance of the two formation channels in a
given globular cluster (GC).  A new approach to solve this problem has
been presented by Mapelli et al. (2004, hereafter paper~I), who
revised and upgraded an original attempt by Sigurdsson, Davies \&
Bolte (1994). The basic input for the new method is the
observation of the shape of the BSS radial distribution in a
star cluster. After the pioneering works on M3 (Ferraro et al. 1993, 1997a) and M55 (Zaggia, Piotto \& Capaccioli 1997), accurate
determinations of the spatial distribution of BSS are now being obtained
for an increasing sample of GCs (Ferraro et al. 2004 for 47~Tuc; Sabbi et
al. 2004 for NGC~6752; Ferraro et al. 2006a for $\omega$~Cen; Warren,
Sandquist \& Bolte 2006 for M5).  A very interesting result is
emerging from these observations: BSS display a clear
tendency to follow a bimodal spatial distribution, with a peak in the
core, decreasing at intermediate radii and rising again at larger
radii.

In paper~I, Mapelli et al. focused on 47~Tuc and showed that the
bimodal distribution can be reproduced assuming a  suitable
combination of the two proposed mechanisms for forming BSS. In
particular, it was suggested (paper~I) that the BSS in the core of
47~Tuc are mainly COL-BSS, while the
external BSS are  MT-BSS.  This result agrees
with studies of the BSS luminosity function in 47~Tuc (Bailyn
\& Pinsonneault 1995; Sills \& Bailyn 1999; Sills et al. 2000; Ferraro
et al. 2003a; Monkman et al. 2006).

In this paper, we explore two other GCs, whose BSS
have a
bimodal radial distribution. The aim is to test whether the conclusions
drawn for 47~Tuc can be extended to GCs having different mass,
concentration, central density, and velocity dispersion with respect
to 47~Tuc. In particular, we study whether the position of a BSS in a GC can
reliably be regarded as a signature for its origin.  We also consider
the case of $\omega$~Cen, whose radial distribution of BSS is flat.
We follow the same procedure of paper~I, carrying out dynamical
simulations of BSS evolved in the gravitational potential well of
their associated cluster. Details about the simulations are
illustrated in Section 2; the sample of the investigated clusters is
presented in Section 3. The results for M3, NGC~6752 and 47~Tuc are
discussed in Section 4, whilst Section 5 is devoted to the
outlier, $\omega$~Cen. In Section 6 we  discuss our findings in comparison with
 the findings of Davies et al. (2004). Finally 
in Section 7 we present our
conclusions.

\section{The simulations}
%%%%%%%%%%%%%%%%%%%%%%%%%%%%%%% TABLE 1%%%%%%%%%%%%%%%%%%%%%%%%%%%%%%%%%
\begin{table*}
\begin{center}
\caption{Globular cluster parameters}
\leavevmode
\begin{tabular}[!h]{llllllc}
\hline
\hline
& $r_{\rm c}$ (pc)
& $\sigma$ (km s$^{-1}$)
& $n_{\rm c}$ (stars pc$^{-3}$)
& $W_0$
& $c$
& references$^{\rm a}$\\
\hline
NGC~104 (47~Tuc) & 0.47 & 10 & $2.5\times{}10^5$ & 12 & 1.95 & 1, 2\\
NGC~5272 (M3) & 1.5 & 4.8 &  $6\times{}10^3$ & 10  &  1.77  & 1, 3\\
NGC~6752 & 0.1, 0.58$^{\rm b}$ & 4.9, 12.4$^{\rm c}$ & $2\times{}10^5$ & 13, 12$^{\rm b}$ & 2.03, 1.95$^{\rm b}$ & 1, 4\\
NGC~5139 ($\omega{}$ Cen) & 4.1 & 17 & $5.6\times{}10^3$ & 6.5 & 1.40 & 5\\
\noalign{\vspace{0.1cm}}
\hline
\end{tabular}
\end{center}
\footnotesize{
$^{\rm {a}}$References:~1~-~Dubath~et~al.~(1997);~2-~Mapelli~et~al.~(2004);~3-~Sigurdsson~et~al.~(1994);~4-~Drukier~et~al.~(2003); 5-~Merritt,~Meylan~\&~Mayor~(1997).

{\hspace{-0.4cm}}$^{\rm {b}}$ The double value of $W_0$ and $c$ for NGC~6752 is due to the
fact that the profile of this cluster is fit with a double
King model. $W_0=13$ and $c=2.03$ refer to the inner King (with core radius
5.7''= 0.1 pc for a distance of 4.3 kpc from the Sun; Ferraro et
al. 2003b); while $W_0=12$ and $c=1.95$ refer to the outer King (with
core radius 28''= 0.58 pc; Ferraro et al. 2003b).

{\hspace{-0.4cm}}$^{\rm {c}}$ The double value of $\sigma{}$ for NGC~6752 refers to the two measurements $\sigma{}$=4.9$^{+2.4}_{-1.4}$ km s$^{-1}$ (Dubath et al. 1997) and $\sigma{}$=12.4$\pm{}0.5$ km s$^{-1}$ (Drukier et al. 2003).
}
\end{table*}
%%%%%%%%%%%%%%%%%%%%%%%%%%%%%%%%%%%%%%%%%%%%%%%%%%%%%%%%%%%%%%%%%%%%%%%%%%%%%
The simulations have been performed with an upgraded version (fully
described in paper~I) of the code originally developed by Sigurdsson
\& Phinney (1995). This code follows the dynamical evolution of a BSS in a
static cluster background, which is modelled using a multi-mass King
density profile. In this case, once the classes of mass have been
selected, the cluster background is uniquely determined by imposing a
central velocity dispersion $\sigma$, a core stellar density
$n_{\rm c}$ and a dimensionless central potential $W_0$
($W_0\equiv\Psi(0)/\langle\sigma\rangle^2$, where
$\langle\sigma\rangle$ is the mean core velocity  dispersion and
$\Psi(0)\equiv\Phi(r_{\rm t})-\Phi(0)$, with $\Phi(r)$ the
gravitational potential at the radius $r$ and $r_{\rm t}$ the tidal
radius). In practice, we have chosen 10 classes of mass (the same
classes reported in table 1 of paper~I) and have adopted the best
available estimates of $\sigma$ and $n_{\rm c}$ for each of the
considered clusters. The values of $W_0$ for each GC have  been
derived by fitting the simulated star density profile to the observed
one (see Section 3 and Table 1 for details).

Once the background has been determined, the dynamical evolution of
the  current 
population of BSS is simulated assuming a value for the ratio 
between the number of MT-BSS and that of COL-BSS.  As far as the birth places
are concerned, we
assume that COL-BSS are generated exclusively in the innermost
region, within the core radius ($r_{\rm c}$),
where the star density is highest, leading to a high collision rate
(Leonard 1989; Pooley et al. 2003).  MT-BSS can be generated everywhere
in the cluster, but are expected to be more frequent in
the peripheral regions, where PBs can evolve in isolation, without
suffering exchange or ionization by gravitational encounters
(Sigurdsson \& Phinney 1993; Ivanova et al. 2005). For this reason, in
our simulations the MT-BSS, formed in PBs, are generated outside the
cluster core with initial locations distributed in several radial
intervals, between 1 and 80 $r_{\rm c}$.  Within this region, all
initial positions are randomly chosen following a flat probability
distribution, according to the fact that the number of stars $N$ in a
King model  %and inside the half mass radius $R_{\rm h}$  
scales as $dN=n(r)\,dV\propto r^{-2}\,\pi r^{2} dr\propto{}dr$.

BSS velocities are randomly generated following the distribution
illustrated in section 3 of Sigurdsson \& Phinney (1995, equation 3.3). In
addition, we assign a natal kick velocity $v_{\rm kick}$ equal to
$1\,{}\sigma$ to those BSS formed collisionally in the core. We
tried also different values of $v_{\rm kick}$. In agreement with paper~I, 
we find that $v_{\rm kick}$ higher than $2-3\,{}\sigma$ causes the
ejection of most of the BSS from the cluster, while there are no
significant differences in the results for kick velocities ranging
between 0 and $2\,{}\sigma$. The masses of the BSS range between 1.2
and 1.5 M$_{\odot}$. In paper~I, we extended our analysis up to a mass
of 2 M$_{\odot}$ (Ferraro et al. 1997a; Gilliland et al. 1998); but now
more stringent constraints are coming from observations (e.g. the upper limit for the mass of BSS in $\omega{}$ Cen is
1.4 M$_{\odot}$; Ferraro et al. 2006a). On the other hand, in paper~I
we found that there is not significant difference between the behaviour
of a 1.2 and a 2 M$_{\odot}$ BSS.  Each single BSS is evolved for a
time $t_i=f_i\,{}t_{\rm last}$, where $f_i$ is a random number uniformly
generated in [0, 1] and $t_{\rm last}$ is the maximum lifetime
attributed to a BSS.
Given the uncertainty on the value of $t_{\rm last},$ we have performed
various sets of runs with $t_{\rm last}$ spanning the range between $1$ and $5$ Gyr. 
Consistent with the results of paper~I, the best fits are all
obtained for $t_{\rm last}$= 1.5 Gyr. For 1 Gyr $\lesssim{}t_{\rm
last}\lesssim{}$ 2 Gyr the distribution does not change
dramatically. If we choose a longer lifetime (3 Gyr or more), the dynamical
friction washes out any peak of peripheral BSS; whereas for shorter
lifetimes (less than 1 Gyr) any BSS could be hardly observable today. 

Our code follows the dynamical evolution of the BSS in the cluster potential,
 subject to the action of  
dynamical friction and to the effects of distant encounters (equation 3.4
of Sigurdsson and Phinney 1995).

The final positions of the BSS in each run determine the
simulated radial distribution, to be compared with the observed one.
We ran 10000 experiments for each of the considered cases. Since we 
studied $\sim{}$10 different cases for each cluster, we made about
$\sim{}$400000 runs in total.

%%%%%%%%%%%%%%%%%%%%%% FIGURE 2%%%%%%%%%%%%%%%%%%%%%%%%%%%%%%%%
\begin{figure*}
  \epsfxsize=15.truecm \epsfysize=15.truecm \epsfbox{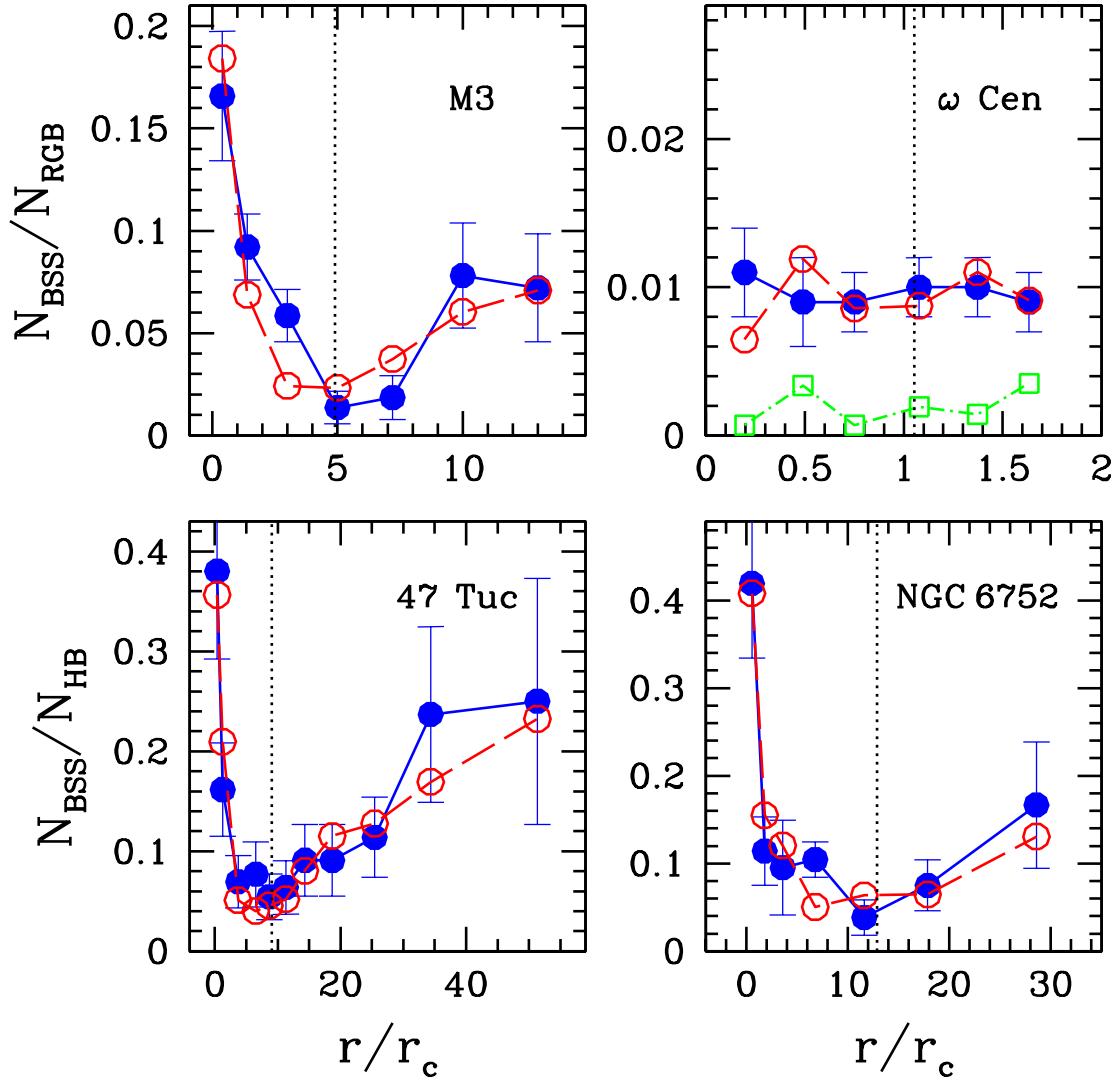}
\label{fig2}
\caption
{Radial distribution of BSS normalized to the distribution of
horizontal branch (HB) stars (47~Tuc, NGC~6752) or red giant branch
(RGB) stars (M3, $\omega{}$ Cen). Filled circles and solid lines
indicate the data taken from Ferraro et al. 1997a (M3),
Ferraro et al. 2006a ($\omega$~Cen), Ferraro et al. 2004 (47~Tuc),
Sabbi et al. 2004 (NGC~6752).  Open circles and dashed lines
indicate the best fits obtained from our simulations. For NGC~6752 we
adopted $r_{\rm c}=28''$ (see caption of Table 1). The dotted vertical
line in all the panels shows the characteristic radius of the zone of
avoidance $r_{\rm av}$ (see text and Table 2). In the panel of
$\omega$~Cen, the open squares and the dot-dashed line connecting them
report the ratio $N_{\rm XMM}/N_{\rm RGB}$, that is the number of
X-ray sources detected in $\omega$~Cen by XMM (Gendre et al. 2003)
normalized to the number of RGB stars.}
\end{figure*}
%%%%%%%%%%%%%%%%%%%%%%%%%%%%%%%%%%%%%%%%%%%%%%%%%%%%%%%%%%%%%%%

\section{The sample of globular clusters}

We have studied the distribution of BSS in four different GCs: M3,
47~Tuc, NGC~6752 and $\omega$~Cen. They represent a very inhomogeneous
sample of GCs. M3 is an intermediate density ($6\times10^3$ stars pc$^{-3}$;
Sigurdsson et al. 1994), low velocity  dispersion (4.8 km
s$^{-1}$; Dubath, Meylan \& Mayor 1997) cluster. 47~Tuc and NGC~6752
are two very concentrated, maybe core collapsed clusters.
$\omega$~Cen is the most massive among the Galactic GCs
($5\times10^6\,{\rm M_\odot}$; Pryor \& Meylan 1993; Meylan et
al. 1995) and it is still a matter of debate whether it is a GC or the remnant
of a dwarf galaxy (Zinnecker et al. 1988; Freeman 1993; Ideta \&
Makino 2004): it is a very loose cluster, with a wide metallicity
spread and a clear evidence for rotation (van Leeuwen et al. 2000;
Reijns et al. 2006; van de Ven et al. 2006).

The first step of our work consisted in finding the best match between
the simulated star density profile of each cluster and the observed
one. The results are shown in Fig.~1. The data of the star density
profile of M3, NGC~6752 and $\omega$~Cen  are from Ferraro et al. (1997a, 2003b) and Ferraro et al. (2006a), respectively.
The star density profile of 47~Tuc is the same as published in paper~I.
 Note that we fit the profile of NGC~6752 with a
double King model, as in Ferraro et al. (2003b). 
 The adopted values of
$\sigma$ and $n_{\rm c}$, and the fit values of $W_0$ are given in Table~1, 
as well as the concentration parameter $c$ (where
$c=\log(r_{\rm t}/r_{\rm c})$) and the fiducial values of $r_{\rm c}$ in
parsec. The latter quantities are derived by assuming a distance from the Sun
of 4.6 kpc (47~Tuc, Ferraro et al. 1999, see also table 3 by 
Beccari et al. 2006), 10.1 kpc (M3, Ferraro et al. 1999), 4.3 kpc
(NGC~6752, Ferraro et al. 1999) and 5.5 kpc ($\omega$~Cen, Bellazzini et
al. 2004).

The data for the radial distribution of BSS are taken from Ferraro et
al. (1997a) for M3, Ferraro et al. (2004) for 47~Tuc, Sabbi et al. (2004) for
NGC~6752 and Ferraro et al. (2006a) for $\omega$~Cen.  As usual, the
number of BSS is normalized to the number of red giant branch (RGB)
stars or to the number of horizontal branch (HB) stars.  Even if the
considered GCs are very different from each other, the observed radial
distributions of BSS are similar (see Fig.~2), with the exception of
$\omega$~Cen. In fact, the radial distribution of BSS displays a maximum in the
central region of the GC, decreases down to a minimum at intermediate
radii (of order $5-10\,r_{\rm c}$) and increases again at outer
radii. $\omega$~Cen differs as it shows an almost flat
distribution. %However, note that the sampled area is limited only to the very central regions (inside $\sim 2r_{\rm c}$). 

\section{Results and Discussion for M3, 47~Tuc and NGC~6752}

We have explored whether the radial distribution of BSS observed in
our sample of GCs can be reproduced with a model which predicts two different
classes of progenitors for the BSS: {\it (i)} PBs evolved in isolation and
{\it (ii)} binary (or single) stars which underwent collisions in the high
density environment of the GC core.  Fig.~2 shows the results obtained for M3,
47~Tuc, NGC~6752 and $\omega$~Cen.  In all cases, our adopted model provides a
statistically acceptable fit to the observations.  The best fit value of the
fraction $\eta{}_{\rm MT}$ of MT-BSS with respect to the total number of BSS [i.e. $\eta{}_{\rm MT}\equiv{}N_{\rm MT-BSS}/(N_{\rm MT-BSS}+N_{\rm COL-BSS})$, where $N_{\rm MT-BSS}$ and $N_{\rm COL-BSS}$ are the number of mass-transfer and collisional BSS, respectively] % resulting from mass transfer in PBs and the number of collisional BSS 
is reported in Table 2, as well as the
reduced $\chi^2$ of the best fit solution.

 In order to determine the best fit solution, we have created a
one-dimensional grid of simulations for each cluster, allowing the
only free parameter in our model\footnote{Other possible parameters
(such as the initial velocity of BSS) were found to have only minor
impact on the results (see Section~2) and have been set to a fixed value in
all the considered runs. Also, the cluster properties are not free
parameters in the fitting procedure, because they have been chosen
{\it a priori} (see Section 3) and kept fixed in all simulations of a given
cluster.~An~exception was made only for the peculiar case of the
velocity dispersion in NGC~6752 (see Section 4.3).}, $\eta{}_{\rm MT}$, to
vary between 0 and 1 with an initial step of $0.2.$ Then, the grid was
iteratively refined down to a step of $0.01$ for the values
surrounding the best fit. 

The values ${\tilde{\chi{}}}^2$ of the
reduced $\chi{}^2$ were derived by comparing each simulation with the
observed data and their uncertainties. The range of values for
$\eta{}_{\rm MT}$ in the second column of Table~2 is calculated at $2\,{}\sigma{}$ level,  i.e. by
selecting all the simulations for which $\chi{}^2\le\chi{}^2_{\rm best}+4,$ where $\chi{}^2_{\rm best}$ is the value for the best fit
solution.

Inspection of Table~2 reveals a clear dichotomy between the cases of 47~Tuc,
 M3 and NGC~6752 and that of $\omega$~Cen. The $2\,{}\sigma{}$ intervals of
 confidence for $\eta{}_{\rm MT}$ for the first three GCs are in agreement
 with an almost even distribution of MT-BSS and COL-BSS (a slight predominance of COL-BSS being indicated by the best fit models). Instead, in the best fit
 model of $\omega$~Cen only 14\% of BSS are COL-BSS and the $2\,{}\sigma{}$
 interval of confidence is compatible with $\eta{}_{\rm MT}=1.$

%All the best fit models indicate a slight predominance of COL-BSS, with the exception of $\omega$~Cen.  In particular, in 47~Tuc, M3 and NGC~6752, the percentage of MT-BSS ranks from 41\% up to 46\%, whereas the percentage of COL-BSS goes from 59\% down to 54\%. Instead, in the best fit simulation of $\omega$~Cen 86\% of BSS are MT-BSS and only 14\% are COL-BSS. Furthermore, the distribution of BSS in $\omega$~Cen is consistent within $2-\sigma{}$ with $\eta{}_{\rm MT}=1$.

In the following, we will discuss the
case of these three clusters separately, whereas $\omega$~Cen will be examined
in Section 5.

\subsection{M3}

In the case of M3, our findings contrast earlier suggestions by
Sigurdsson et al. (1994). They hypothesized that the BSS
observed in the peripheral regions of M3 could be COL-BSS
born initially in the core and then ejected in the periphery because
of their natal kick. Indeed, our dynamical simulations show that
COL-BSS are ejected from the entire cluster (if the kick is
too high), or sink back to the core in $\simlt 1$ Gyr, if the kick is
low enough to retain them in the cluster potential well. Hence, they
cannot account for the peripheral BSS. Instead, MT-BSS
have in general much more circular orbits than the
COL-BSS and can remain in the peripheral regions
of the GC for a time comparable to the typical lifetime of a BSS.

\subsection{47~Tuc}

As for 47~Tuc, we confirm the main outcome from paper~I:
 i.e. most of the central BSS of 47~Tuc are born from collisions and all the
peripheral BSS originate from mass-transfer in binaries. However, in paper~I
it was suggested that only $\sim25$\% of the total BSS are MT-BSS.
The upward correction of this estimate is due to the fact that we have
refined our parameter grid with respect to paper~I. In paper~I we considered
only cases with a relatively low fraction of MT-BSS 
($\lesssim{}30$\%) and one single case with $\eta{}_{\rm MT}=1$. For this paper
we explored a more complete parameter space, the percentage of MT-BSS (with respect to the total number of BSS) going from 0 to 100\%. Our present best fit (46\% MT-BSS and 54\% COL-BSS) is still consistent (at $2\,{}\sigma{}$) with the result of paper~I (25\%
MT-BSS and 75\% COL-BSS), given the observational
uncertainties, especially in the central bins (see Fig.~2).

\subsection{NGC~6752}

The parameters of the best fit model for NGC~6752 are similar to those
inferred for 
M3 and 47~Tuc.  However, for this cluster there may be a
caveat about the central velocity dispersion, whose value is still
very uncertain. From the integrated-light spectra of the core, Dubath
et al. (1997) derived $\sigma=4.9^{+2.4}_{-1.4}$ km s$^{-1}$. More
recently, from proper motion measurements of a sample of $\sim1000$
stars with the Wide Field Planetary Camera 2/{\it Hubble Space Telescope} (WFPC2/HST), Drukier et al. (2003) obtained
$\sigma=12.4\pm0.5$ km s$^{-1}$. Thus, we ran simulations for each of these two
possibilities.  
%{\bf ANDREA: IS NOT THERE ALSO AN ESTIMATE BY SOME PUPIL OF
%PRYOR?  I REMEMBER SOMETHING ABOUT 7 km/s. WAS THAT EVENTUALLY
%PUBLISHED? MICHELA: NON SO. FRANCESCO, NE SAI QUALCOSA?}

If we assume $t_{\rm last}=1.5$ Gyr, the best fit model obtained for $\sigma=4.9$ km s$^{-1}$ (Fig.~3) is not
satisfactory, since we cannot reproduce the increasing number of BSS located in
the cluster outskirts. In fact, a low velocity dispersion  implies that the
dynamical friction time-scale is too short ($\ll{}1$ Gyr) and PBs rapidly
shrink towards the core.
%escape velocity from the cluster halo is too low to retain a sizeable fraction of the external BSS.  
If we adopt $\sigma=12.4$ km s$^{-1}$, the rising trend of the BSS radial
distribution in the cluster periphery can be reproduced; but the 
$\tilde{\chi{}}^2$ ($\sim{}5$) is still unacceptable, because of the disagreement 
with the data
points in the inner regions.  Since none of the two options was satisfying,
we ran new simulations for other values of $\sigma$ (ranging between 4.9 and
12.4 km s$^{-1}$), until we found an acceptable fit to the observed BSS
distribution.  That occurs for $\sigma=7^{+3}_{-1}$ km s$^{-1}$ ($\tilde{\chi{}}^2=1.7$, reported in Table~2), marginally consistent with the
measurement of Dubath et al. (1997).

 However, for this cluster other marginally acceptable solutions for
the BSS radial profile exist: e.g. assuming $t_{\rm last}=4$ Gyr, a
fit ($\tilde{\chi{}}^2\lesssim{}3$) can be obtained also for
$\sigma=12.4$ km s$^{-1}$ (open triangles and short-dashed line in
Fig.~3).  Considering  this uncertainty and the peculiarities of NGC~6752 (e.g. the
unusual density profile), the aforementioned best fit value
$\sigma=7^{+3}_{-1}$ km s$^{-1}$ cannot be taken as an indirect
measurement of the central velocity dispersion in the cluster. A
new measurement of $\sigma$ is strongly needed
to understand the dynamics and evolution of NGC~6752.

%Given these findings, a new determination of the core 
%velocity  dispersion of NGC~6752 is crucial.  If $\sim{}7$ km s$^{-1}$ will result a
%favored value for $\sigma$, that would strongly support our modeling
%of the formation and dynamics of BSS in a GC.

%%%%%%%%%%%%%%%%%%%%%% FIGURE 3%%%%%%%%%%%%%%%%%%%%%%%%%%%%%%%%
\begin{figure} 
\epsfxsize=8.truecm 
\epsfysize=8.truecm
\epsfbox{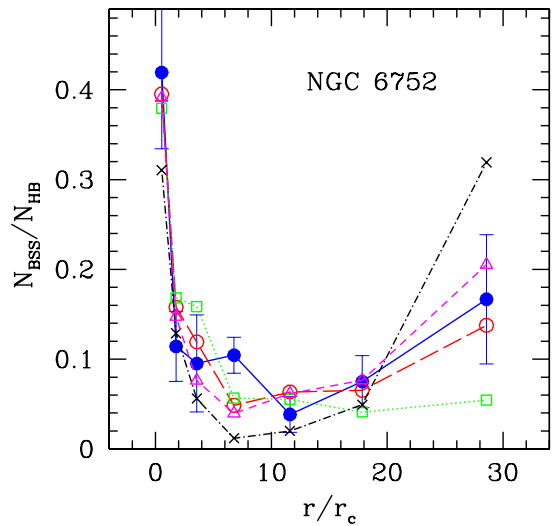}
\label{fig3}
\vspace{-0.5truecm}
\caption
{Distribution of BSS normalized to the distribution of
HB stars in NGC~6752. The solid line
indicates the observations (filled circles; Sabbi et al. 2004). The dotted
line (connecting open squares) reports the best fit numerical
simulation for $\sigma=4.9$ km s$^{-1}$; the long-dashed line
(connecting  open circles) is for $\sigma=7$ km s$^{-1}$ (reported also in Fig.
~2); the
 dot-dashed line (connecting crosses) is for the case
$\sigma=12.4$ km s$^{-1}$. In all these cases $t_{\rm last}=1.5$ Gyr. The short-dashed line connecting open triangles is obtained for $\sigma=12.4$ km s$^{-1}$ and $t_{\rm last}=4$ Gyr.}
\end{figure}
%%%%%%%%%%%%%%%%%%%%%%%%%%%%%%%%%%%%%%%%%%%%%%%%%%%%%%%%%%%%%%%

\subsection{The location of the minimum in the BSS radial distributions}

The radial distribution of BSS (normalized to the number of RGB or HB
stars) in M3, 47~Tuc, and NGC~6752 clearly displays a minimum at a
distance from the cluster center in the range $5-10\,{}r_{\rm c}.$ The
area surrounding this minimum was dubbed 'zone of avoidance' (paper~I).  
We expect that its location roughly corresponds to the radius
$r_{\rm av}$ below which all the PBs of mass $\simgt
m_{\rm BSS}$ (where $m_{\rm BSS}$ is the minimum mass for generating BSS
via mass transfer) have already sunk towards the cluster center
because of dynamical friction. In other words, there is a small
probability of observing a MT-BSS
located at a distance from the cluster center of the order of $r_{\rm
av}.$ Most likely, the MT-BSS  which were
originally located at $r\simlt r_{\rm av}$ can now be detected at
$r\simlt r_{\rm c}$.

The time of dynamical friction $t_{\rm df}$ for a mass $m_{\rm BSS}$
located at a radius $r$ from the center of a GC is given by (Binney \&
Tremaine 1987)
\q
\label{eq:eq1}
t_{\rm df}=\frac{3}{4\,\ln{\Lambda}\,G^2(2\,\pi)^{1/2}}\,
\frac{\sigma(r)^3}{m_{\rm BSS}\,\rho(r)}, \nq where $\ln{\Lambda}\sim10$ is the
Coulomb logarithm and $G$ the gravitational constant. $\rho(r)$ and $\sigma(r)$ are the cluster
density and the velocity dispersion at the radius $r$. 
We note the dependence of $t_{\rm df}$ on the density $\rho(r)$ and especially on the cube
of the velocity dispersion. Given the definition of this time-scale, $r_{\rm av}$
can be calculated setting $t_{\rm df}=t_{\rm gc}$, where $t_{\rm gc}$ is the
lifetime of the cluster ($\sim{}$12 Gyr), and using the simulated stellar density profile and the velocity dispersion distribution of
the cluster for inferring $r_{\rm av}$ from $\rho(r_{\rm av})$ and $\sigma(r_{\rm av})$. In doing the
calculation, we choose $m_{\rm BSS}=1.2\,{\rm M_\odot}.$

The results (for $r_{\rm av}$ normalized to $r_{\rm c}$) are shown in
the last column of Table 2 and in Fig.~2 (vertical dotted line). The
values of $r_{\rm av}$ are always consistent with the zone of avoidance 
observed in the data. This
confirms for M3 and NGC~6752 what had been found for 47~Tuc (paper~I).

Third and fourth columns of Table~2 report the values of the fraction ${\eta{}}_{\rm MT}$ {\it  (i)} for the MT-BSS located within the GC core $r_{\rm c}$ ($\eta{}_{\rm MT} (<r_{\rm c})$), and  {\it (ii)} for the MT-BSS found at a distance from the cluster center larger than $r_{\rm av}$ ($\eta{}_{\rm MT} (>r_{\rm av})$).  These findings
indicate that the position of a BSS in a GC (at least in those
showing a bimodal distribution of the BSS) can be used as a strong indication
of the nature of the progenitor. In particular, a BSS which is located outside
the zone of avoidance almost certainly is a MT-MSS;
%, i.e. results from the evolution of a PB; 
on the contrary, a BSS found close to the cluster core is a COL-BSS.
%, i.e. has a collisional origin.  

 Among the considered clusters,  only $\omega$ Cen does not show any zone of avoidance. If we calculate the expected radius of avoidance also for this cluster, we find $r_{\rm av}\sim{}r_{\rm c}$. This indicates that in $\omega$ Cen the dynamical friction is far less efficient than in the other clusters.
In $\omega{}$~Cen the value of ${\eta{}}_{\rm MT}(>r_{\rm av})$ is similar to the other clusters, whereas ${\eta{}}_{\rm MT}(<r_{\rm c})$ is much higher (see Section 5).

We note that very similar values of both $\eta{}_{\rm MT}$ and $r_{\rm av}/r_{\rm c}$
have been obtained for an intermediate density GC (M3) and for two much denser clusters (47~Tuc and NGC~6752). The analogy between
these three clusters is even more evident in Fig.~4, where radial
distances are normalized to the avoidance radius, $r_{\rm av}$. The radial distributions of BSS in M3, 47~Tuc and NGC~6752 nearly superimpose.  This also indicates that $r_{\rm av}$ is a crucial parameter in describing the bimodal distribution of BSS.

%%%%%%%%%%%%%%%%%%%%%% FIGURE 4%%%%%%%%%%%%%%%%%%%%%%%%%%%%%%%%
{ 
\begin{figure}{ 
\center{{
\epsfxsize=8.truecm \epsfysize=8.truecm
\epsfbox{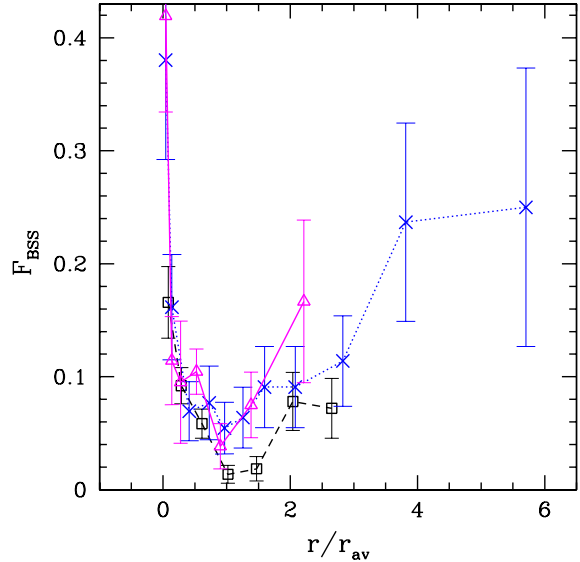}
}}\label{fig4}
\footnotesize 
\caption{
  Observed distribution of BSS normalized to the distribution of HB (47~Tuc,
  NGC~6752) or RGB stars (M3). On the x-axis we plot the
  distance from the center normalized to $r_{\rm av}$. Crosses indicate 
  the observational data for 47~Tuc, open squares for M3 and open triangles for NGC~6752. The
  observational points are connected by a dashed line for M3, a dotted line for 47~Tuc and a solid line for
  NGC~6752.  }

}
\end{figure}

%%%%%%%%%%%%%%%%%%%%%%%%%%%%%%% TABLE 2%%%%%%%%%%%%%%%%%%%%%%%%
\begin{table}
\begin{center}
\tabcolsep 4pt
\leavevmode
\caption{Mass Transfer vs Collisional Scenario}
\begin{tabular}{lccccc}
\hline
\hline
& ${\eta{}}_{\rm MT}$ 
& ${\eta{}}_{\rm MT}(<r_{\rm c})$ 
& ${\eta{}}_{\rm MT}(>r_{\rm av})$  
& $\tilde{\chi}^2$
& $r_{\rm av}/r_{\rm c}$\\
\hline
NGC~104       & 0.46$^{+0.06}_{-0.22}$ & 0$^{+0.002}_{-0}$  & 0.95$^{+0.02}_{-0.10}$    &  0.5  &  9.0\vspace{0.1cm}\\
NGC~5272      & 0.41$^{+0.16}_{-0.20}$ & 0$^{+0.002}_{-0}$  & 0.91$^{+0.02}_{-0.01}$     &  2.2  &  4.9\vspace{0.1cm}\\
NGC~6752      & 0.41$^{+0.12}_{-0.04}$ & 0.02$^{+0.03}_{-0.02}$ & 0.97$^{+0.01}_{-0.05}$  &  1.7  & 12.9\vspace{0.1cm}\\
NGC~5139      & 0.86$^{+0.14}_{-0.08}$ & 0.64$^{+0.36}_{-0.04}$  & 0.90$^{+0.10}_{-0.002}$ &  0.8  &  1.05\vspace{0.1cm}\\
%NGC~104 (47~Tuc)        & 0.46 (0.54) & 0 (1) & 0.96 (0.04) & 0.5  & 10.0\vspace{0.1cm}\\
%NGC~5272 (M3)           & 0.41 (0.59) &  5$\times{}10^{-4}$ (0.9995) & 0.90 (0.10) & 2.2 &  4.6\vspace{0.1cm}\\
%NGC~6752                & 0.41 (0.59) & 0.02 (0.98) & 0.975 (0.025) & 1.7 & 12.9\vspace{0.1cm}\\
%NGC~5139 ($\omega$ Cen) & 0.86 (0.14) & 0.64 (0.35) & 0.90  (0.10)  &  0.8 &  1.0\vspace{0.1cm}\\
%NGC~104 (47~Tuc)        & 0.86$_{-0.6}^{+0.2}$ & 22.8 & 0 &  0.5  & 10.0\vspace{0.1cm}\\
%NGC~5272 (M3)           & 0.69$_{-0.5}^{+0.4}$ &  9.4 & 5$\times{}$10$^{-4}$  & 2.2 &  4.6\vspace{0.1cm}\\
%NGC~6752                & 0.69$_{-0.1}^{+0.3}$ & 38.2 & 0.02 & 1.7 & 12.9\vspace{0.1cm}\\
%NGC~5139 ($\omega$ Cen) & 6.3$_{-3}^{+4}$ &  8.8 & 1.8  &  0.8 &  1.0\vspace{0.1cm}\\
\noalign{\vspace{0.1cm}}
\hline
\end{tabular}
\end{center}
\footnotesize
{
The second column reports the fraction $\eta{}_{\rm MT}$ of MT-BSS with respect 
to the total number of BSS (the errors are at 2$\,{}\sigma{}$, i.e. referred to $\chi{}^2=\chi{}_{\rm best}^2+4$). The third (fourth) column reports
the fraction of MT-BSS inside $r_{\rm c}$ (outside  $r_{\rm av}$) with respect
 to the total number of BSS inside $r_{\rm c}$ (outside  $r_{\rm av}$).
%In particular, ''collisional BSS'' (''mass-transfer BSS'') are all those BSS born within (beyond) 1 core radius and with (without) a kick velocity of $1\times\sigma$. 
The fifth column lists the reduced
$\chi^2$ obtained for the best fit models. The rightmost column
reports $r_{\rm av}$ normalized to $r_{\rm c}.$ 
}
\end{table}
%%%%%%%%%%%%%%%%%%%%%%%%%%%%%%%%%%%%%%%%%%%%%%%%%%%%%%%%%%%%%%%

\section{The case of $\omega$ Centauri}

The main characteristic of the BSS radial distribution of $\omega$~Cen,
with respect to the other clusters discussed here, is the absence of 
any central peak. In Fig.~2 
we show the high-resolution portion of the data set presented by
Ferraro et al. (2006a): the distribution appears to be flat. This is
quite unusual, since   in all the other GCs surveyed
up to now the BSS distribution is peaked at the cluster center. 
Fig.~6 by Ferraro et al. (2006a) suggests that this behaviour
extends up to at least $\sim{}7\,r_{\rm c}$ ($\sim 20'$ from the cluster center).

This peculiarity adds to the many other unique features displayed by
$\omega$~Cen (see Section 3); but our simple model provides a satisfactory
fit to the data (see Fig.~2 and Table~2) also in this case.
However, at variance with the results for the three clusters examined
in Section 4, the best fit model requires that a large majority of the BSS
are born from PBs (see Table~2). In other words, only a mere $\sim{}
14\%$ of the BSS detected in  $\omega$~Cen may have a
collisional origin, i.e. we expect that only $\sim{}44$ of the 
313 BSS reported by Ferraro et al. (2006a) are COL-BSS.

The alternative hypothesis that all the others BSS formed via
collisions have been ejected from $\omega$~Cen sounds unrealistic,
since the central escape velocity of $\omega$~Cen ($\sim 40$ km
s$^{-1}$) is comparable with that of other clusters in our sample and
there is no motivation for assuming that collisions in $\omega$~Cen
should generate larger recoil velocities with respect to other
clusters. As a consequence, the dynamical modelling of the BSS
presented in this work predicts that COL-BSS are now produced
with a low rate in $\omega$~Cen.

Are there viable explanations for this underproduction of COL-BSS
in $\omega$ Cen?
We can just note that the characteristic radius $r_{\rm av}$ for
$\omega$~Cen is much smaller (both in terms of $r_{\rm c}$ and of
physical units) than in the other clusters of our sample ($r_{\rm
av}\sim{}\,r_{\rm c}$ instead of the typical value ranging between 
$5~r_{\rm c}$ and $15~r_{\rm c}$). This means that only
a small number of PBs had enough time to sink into the
core due to mass segregation. 

PBs  that sink within the core are thought to be the progenitors  (as 
far as modified by three body interactions) of most of the  binaries hosted in 
the core of current GCs, since the formation of binaries 
from two-body interactions is a quite unlikely process.
Because of their large cross-section, core binaries 
suffer repeated three- and four-body
interactions, eventually exchanging companions, being ionized or hardened.
Thus, core binaries are required  to form COL-BSS, via three- or 
four-body encounters.
Then, the relative lack of binaries in the core of $\omega{}$ Cen, due to the 
inefficiency of dynamical friction, could have quenched the formation of 
COL-BSS.

This view could be strongly supported by the observation of a
low ($\ll{}10$ \%) fraction of binaries 
in the core of $\omega$~Cen.  Unfortunately, this fraction cannot be
derived with classical methods (Rubenstein \& Bailyn 1997; Bellazzini
et al. 2002), because of the wide spread in metallicity of the stars
in this GC (Norris, Freeman \& Mighell 1996; Suntzeff \& Kraft 1996).
An indirect indication of the occurrence of a low fraction of binaries in
the core of $\omega$~Cen
comes from inspection of Fig.~2, which shows the radial
distribution of the X-ray sources observed in $\omega$~Cen by XMM
(Gendre, Barret \& Webb 2003), normalized to the number of RGB stars. Most of
the X-ray sources are believed to be cataclysmic variables or low mass
X-ray binaries (Gendre et al. 2003). So they should be among the most massive
objects in the GC. Their flat radial distribution further supports the
hypothesis that mass segregation in $\omega$~Cen has not driven yet a
sizeable number of massive binaries in the central region.

%\section{The number of BSS and the absolute magnitude of the host GC}
\section{Number of BSS versus $M_V$}
Until now, we only considered the relative frequency of BSS, i.e. the 
number of BSS normalized to that of HB and/or RGB stars. But what happens 
if we consider just the number of BSS hosted in each cluster, without any 
normalization?

First of all, it is worth noting that NGC~6752 ($\omega$ Cen) hosts about three times fewer (more) BSS than either 47~Tuc or M3 (Fig. 5). We can ask whether this is at all
correlated to the mass of the host cluster, or, equivalently, to its absolute magnitude $M_V$? 
Plotting in Fig.~5 the total number of observed BSS per cluster against $M_V$
 for our four clusters, we do not see any straight correlation. This result agrees with the fact that Piotto et al. (2004) found evidence, in a large sample of GCs, of a lack of correlation of the number of observed BSS with either stellar collision rate (as would be expected if all BSS were collisional in origin) or mass (which is proportional to the collision rate). %XXX MICHELA DO YOU MEAN THAT? OR DO YOU MEAN M-T RATE??? 

This fact led  Davies et al. (2004) to propose  that BSS are made through both channels, suggesting that the number of COL-BSS  tends to increase with cluster mass, while MT-BSS  tend to decrease with total mass, as PBs drifting by dynamical friction in the core had 
already time to burn as BSS in the past. The combination of these two
competing behaviours was found to  produce a population of BSS weakly dependent on the total mass and core collision rate of their host GC. 

Fig.~5 sketches the predictions by Davies et al. (2004). In particular, the 
dashed (dotted) line indicates the contribution expected from the MT- (COL-)BSS; the solid line is the total.

47~Tuc and M3 lie close to the solid line and, according to the Davies et al.'s model,
 their current BSS population is obtained by a blending of COL-BSS and 
MT-BSS
contributing in roughly equal number.  This fully 
agrees with what we found from the comparison between 
our simulations and the radial distribution of BSS.

 In contrast, our model and Davies et al.'s scenario disagree in the case of both NGC~6752 and $\omega{}$~Cen. NGC~6752 lies well below  the solid curve. According to Davies et al. (2004), its BSS should  predominantly be MT-BSS; whereas in our model NGC~6752 should host MT-BSS and COL-BSS nearly in equal number.
In the Davies et al.'s model $\omega{}$~Cen is expected to have only COL-BSS, while we have shown that MT-BSS should dominate\footnote{Davies et al. (2004) admit that their model is not adaptable to slow evolving GCs like $\omega$~Cen.}.

Moreover, though the picture proposed by Davies et al. (2004) 
is interesting, detailed cluster-to-cluster comparisons suggest a much more complex
scenario, where the dynamical history, the original 
PB content and the current dynamical state of each
cluster seem to play a major role (Ferraro et al. 2003a).
Indeed,
clusters with the same integrated magnitude harbor quite different BSS
populations. In particular, two GC pairs offer the possibility of demonstrating the
complexity of the emerging scenario:
{\it (i)} M3 and M13 are almost twins GCs (Ferraro et al. 1997b). They have the same 
integrated magnitude ($M_V\sim -8.6$), same metallicity ($[{\rm Fe}/{\rm H}]=-1.6$), same mass,
but display a quite different BSS content (Ferraro et al. 2003a). In 
particular, M13 harbors a factor 5 fewer  BSS than M3. 
%Which is the origin of
%the paucity of BSS in M13?  A number of hypothesis
%which have to be confirmed (difference in the PB content, different evolutionary stage)
%have been suggested (see Ferraro et al. 2003a).
{\it (ii)} NGC~6752 has an integrated magnitude ($M_V\sim -7.7$) which is quite 
similar to M80 ($M_V\sim -7.9$); but again the BSS content in the two clusters
 turns out to be quite different: NGC~6752 harbors a BSS population 
 which is a factor 10 lower than that found in M80
(note that the BSS population of M80 is comparable in size to that found in $\omega{}$ Cen).
Ferraro et al. (1999) suggested that the anomalously large population of BSS in M80
could have originated in the core-collapsing phase of this cluster.

 The model discussed in this paper is based on a different approach
from that of Davies et al. (2004). Neither this work nor Davies et
al.'s accounts for the dynamical evolution of the
cluster; but we consider
additional information in comparison with Davies et al. (2004),
 i.e. the observed BSS radial distribution.

Davies et al. (2004) try to predict the properties of the
BSS from the present properties of the host cluster, and these
properties might not reflect all the stages in the cluster evolution. In
particular, they risk overlooking the importance of the earlier evolutionary phases which
can have a strong impact on the characteristics of  BSS
population. 
%which formed much later and is visible today.

In our approach the cluster evolution is frozen for the last 2 Gyr
only, when the cluster properties are not expected to have changed
significantly. All the effects of the previous evolution of the 
cluster onto BSS  are intrinsically
stored in the initial parameters which we impose on the BSS population:
masses, lifetimes, velocities, locations and the amount of injected
MT-BSS and COL-BSS. 
%Of course, the subsequent evolution of the simulated population is strongly affected by the cluster (fixed) parameters. However, we assume that also the typical masses, lifetimes and velocities of the BSS population do not vary significantly during the last 2 Gyr and in Section~2 it has been shown that the allowed range for these three quantities is indeed relatively small.  
Hence, we can use the observed BSS radial profile for
inferring the ratio between the injected MT-BSS and COL-BSS (which is
related to their location), and for investigating the effects of the
cluster (present) properties on that (e.g. what is the importance of
collisions in the recent life of the cluster; and, what is the residual
fraction of PBs needed to remain in periphery up to now).

Both these methods need refinements; however, the approach presented 
in this paper has the advantage that it is able to fit
the BSS population also in clusters where Davies et al.'s  method
appears inaccurate. A complete understanding of the relation between
a cluster and its BSS population will likely be possible only by running
N-body simulations of the cluster from its formation, accounting also
for three-body encounters, PB evolution, stellar evolution, etc.

%%%%%%%%%%%%%%%%%%%%%% FIGURE 5%%%%%%%%%%%%%%%%%%%%%%%%%%%%%%%%
\begin{figure} 
\center{{
\epsfxsize=8.truecm \epsfysize=8.truecm
\epsfbox{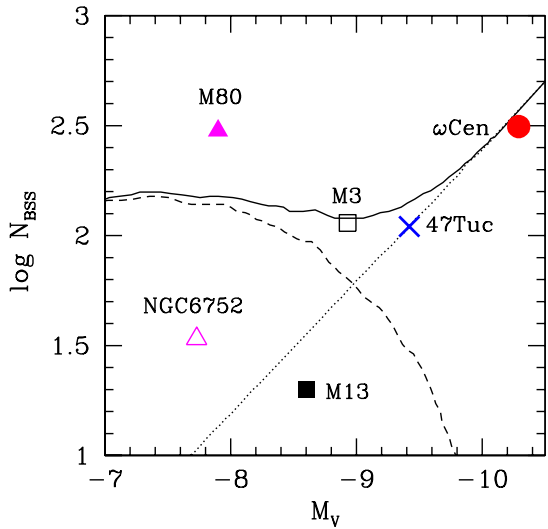}
}}\label{fig5} 
\vspace{-0.5truecm}
\caption{Number of BSS as a function of the cluster
  absolute magnitude $M_V$. Open triangle refers to NGC~6752, filled triangle to M80,
  open square to M3, filled square to M13, cross to 47~Tuc, filled circle to
  $\omega$~Cen. The values of $M_V$ come from the Harris catalogue (Harris 1996; {\tt http://physwww.mcmaster.ca/\%{}7Eharris/mwgc.dat}).
The solid line shows the model by Davies et al. (2004), while the dashed (dotted) line refers  to the MT-BSS (COL-BSS) according to the same model.}
\end{figure}
%%%%%%%%%%%%%%%%%%%%%%%%%%%%%%%%%%%%%%%%%%%%%%%%%%%%%%%%%%%%%%%%%%%%%%%%

\section{Summary}

We have exploited the recent determination of the radial distribution of BSS
in four GCs, in order to investigate which mechanism of BSS formation
prevails in these stellar systems.  Our conclusion is that the two main
formation paths proposed so far,  i.e. mass-transfer in PBs and merging of MS 
stars due to collisions in the cluster core,
must {\it co-exist and have similar efficiency both in a low density cluster
(M3) and in much denser clusters, like 47~Tuc and NGC~6752.}

In particular, in M3, 47~Tuc, and NGC~6752 the COL-BSS 
sum to $\sim 50-60\%$ of the total and mostly
reside in the central region of the cluster. The MT-BSS are slightly less 
abundant than the COL-BSS, but
populate all the GC.  The density of BSS reaches a minimum in
a so-called zone of avoidance, which separates the portion of the GC
mostly occupied by COL-BSS from the cluster outskirts, where
the MT-BSS dominate.  The location of the
zone of avoidance is explained by accounting for the effects 
of the dynamical friction on the PBs which were
massive enough for generating the observed BSS.

The picture described above can also be applied to $\omega$~Cen; but in this
case the lack of a central peak in the BSS radial distribution requires that
the 
large majority of the BSS derive from PBs. The very low rate of production of
COL-BSS could be in turn attributed to the fact that mass segregation
has not yet driven a sizeable number of PBs to the central region of the
cluster to produce BSS.

A very interesting further development of this research will be to perform a
comparison between the location of a significant sample of BSS in a GC and
their spectroscopic properties. According to the findings of this work, the
position in the GC might represent a strong dynamical clue for the formation
mechanism of a given BSS. If it is located outside the zone of avoidance, the
BSS almost certainly results from evolution of a PB; if it is harbored in the
cluster core, the BSS has most likely a collisional origin.  On the other hand,
indication about the origin of the same BSS can be independently obtained from
high resolution spectroscopy. Indeed the chemical signature of the MT-BSS formation process has been recently discovered in 47~Tuc (Ferraro et al. 2006b).
The acquisition of similar sets of data in clusters with different structural parameters
and/or in different regions of the same cluster  
 will provide an unprecedented tool for confirming the scenario presented here
 and to finally address the BSS formation
 processes and their complex interplay with the dynamical evolution
of the cluster.

\section {Acknowledgments}
We thank Craig Heinke, Fred Rasio and Emanuele Ripamonti for useful discussions. We acknowledge the anonymous referee for the critical reading of the manuscript. 
M.~M. acknowledges
 the Northwestern University and the Center for Gravitational Wave Physics 
(Pennsylvania State University) for kind hospitality. M.~C. and A.~P. acknowledge grant 
PRIN05 2005024090\_{}002.
This research was supported by the contract ASI-INAF I/023/05/0.

\end{document}